\documentclass[prl,twocolumn,showpacs]{revtex4}

\usepackage{graphicx}
\usepackage{dcolumn}
\usepackage{amsmath}

\begin{document}

\title{Indication of a non-magnetic surface layer on a magnetic single crystal}

\author{Daniel Wegner}
\email{wegner@berkeley.edu} \altaffiliation[Current address:
]{Department of Physics, University of California at Berkeley,
USA} \affiliation{Institut f{\"u}r Experimentalphysik, Freie
Universit{\"a}t Berlin, Arnimallee 14, 14195 Berlin-Dahlem,
Germany}

\author{G{\"u}nter Kaindl}
\email{kaindl@physik.fu-berlin.de}
\affiliation{Institut f{\"u}r Experimentalphysik, Freie Universit{\"a}t Berlin, Arnimallee 14, 14195 Berlin-Dahlem, Germany}

\date{March 27, 2009}

\begin{abstract}

The structural and electronic properties of the surfaces of
Sm(0001) and Eu/Gd(0001) were studied by scanning tunneling
microscopy and spectroscopy at temperatures between 10 and 110~K.
In both systems, an unoccupied surface state is observed that
exhibits a temperature-dependent splitting into two states for
Eu/Gd(0001), while it is unsplit on Sm(0001). This strongly
indicates that the divalent outermost surface layer of Sm(0001) is
non-magnetic despite the antiferromagnetic trivalent Sm substrate.
These findings open new opportunities for magnetic studies of
ultra-thin Sm films.

\end{abstract}


\maketitle

%
%

The magnetic properties of trivalent lanthanide (Ln) metal
surfaces have been the subject of some controversy in the past.
For Gd(0001), e.g., the possibility of an enhanced surface Curie
temperature ($T_C$) and an orientation of surface magnetization
different from the bulk has been debated \cite{Dow97}. Some of
this behavior could be due to the magnetic exchange splitting of
surface states that does not follow a simple Stoner-like behavior
but is also influenced by spin mixing and short-range magnetic
order at higher temperatures $T$ \cite{Bod99,Weg06c}. More recent
experiments have convincingly shown that the magnetic properties
of most Ln-metal surfaces are essentially the same as in the bulk
\cite{Arn00,Mai02}.

The situation is quite different for Sm metal, where a strong
deviation of surface from bulk magnetism can be expected due to
the well-known surface valence transition \cite{Wer78}. While Sm
in the bulk is trivalent and magnetic with an $^{6}H_{5/2}$ ground
state $[4f^5(6s5d)^3]$, the reduced coordination at the surface
leads to a valence change to divalent Sm with a non-magnetic
$^{7}F_0$ ground state $[4f^6(6s5d)^2]$ \cite{Wer78,Joh79}. This
valence change is accompanied by a dramatic increase of the ionic
radius by 22~\% as well as a reconstruction of the Sm(0001)
surface \cite{Ste89,Lun02}. Since Sm metal orders
antiferromagnetically (AFM) in the bulk below $T_N = 106\,
\text{K}$ \cite{Koe72}, the question arises whether the outermost
Sm(0001) surface layer is magnetic or non-magnetic below $T_N$.

The present study shows that Sm(0001) exhibits a previously
unnoticed \emph{d}-like surface state that allows to address this
question by studying its magnetic exchange splitting via scanning
tunneling spectroscopy (STS) as a function of $T$ ($10\, \text{K}
\leq T \leq 110\, \text{K}$). For comparison, a monolayer (ML) of
Eu on Gd(0001) was studied that exhibits an analogous surface
reconstruction but differs in its local $4f$ moments: both Eu and
Gd have non-vanishing $4f$ moments ($S = J = 7/2$) and couple
ferromagnetically (FM) below $T_C = 293\, \text{K}$
\cite{Are98,Ded06}. The outermost divalent Eu layer exhibits also
a \emph{d}-like surface state. While we observe a $T$-dependent
splitting of the Eu/Gd(0001) surface state, the analogous surface
state on Sm(0001) consists of a single peak. This strongly
indicates that the outermost surface layer of Sm metal is
non-magnetic.

%
%

The experiments were performed in ultrahigh vacuum (base pressure
$< 3 \times 10^{-11}\, \text{mbar}$) with a home-built low-$T$ STM
operated between 10 and 110~K \cite{Bau02}. All Ln-metal films
were deposited \emph{in situ} by electron-beam evaporation of
99.99\% pure metals from a Ta crucible onto a clean W(110) single
crystal kept at room temperature (RT). The 10-ML-thick Sm films
were not annealed upon deposition, since RT deposition readily
leads to smooth films \cite{Lun02}. For Eu/Gd(0001), first a
30-ML-thick Gd(0001) film was grown on W(110) and annealed to
obtain a smooth, crystalline film \cite{Reh03}, followed by
deposition of about 1~ML Eu on the Gd film kept at RT. The samples
were then transferred in UHV to the cryogenic STM. STS spectra
were recorded with fixed tip position and switched-off feedback
loop using standard lock-in techniques (modulation amplitude: 1~mV
(rms), modulation frequency: $\simeq 360\, \text{Hz}$). As is well
known, the differential conductivity, $dI/dV$, is approximately
proportional to the local density of states of the surface ($I =$
tunneling current; $V =$ sample bias voltage).

%
%

\begin{figure}
\begin{center}
\includegraphics{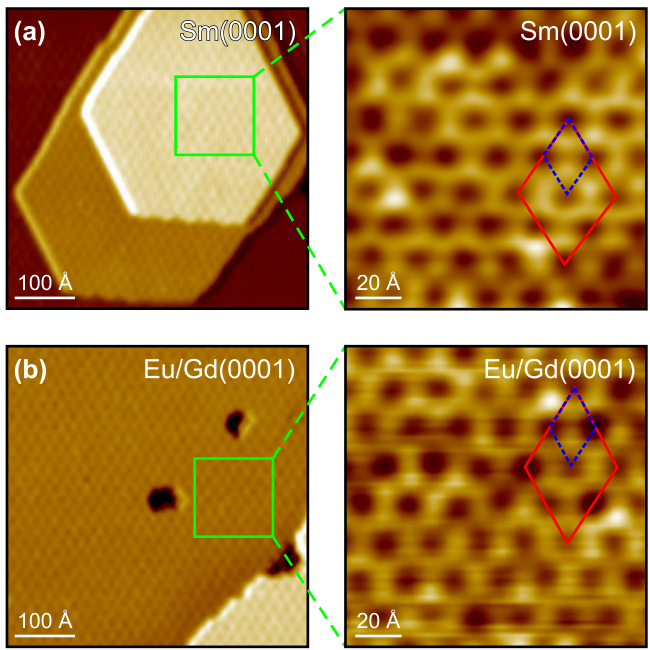}
\caption{\label{fig1} (Color online) (a)~STM images of 10~ML
Sm(0001)/W(110). The close-up image reveals the hexagonal Moir\'e
pattern of the ($11 \times 11$) surface reconstruction (large red
diamond), with an effective ($5.5 \times 5.5$) unit cell in STM
topography (dashed blue diamond). (b)~STM images of 1~ML
Eu/Gd(0001) reveal striking structural similarities. Small holes
in the surface layer indicate that the amount of deposited Eu is
slightly less than 1~ML.}
\end{center}
\end{figure}

Fig.~\ref{fig1}(a) shows the topography of a 10-ML Sm(0001) film.
The surface is atomically flat with ML-high terraces. The close-up
view on a terrace shows the hexagonal Moir\'e pattern of the
well-known surface reconstruction caused by the larger radius of
divalent Sm surface atoms. The structure was first identified  by
low-energy electron diffraction as a $(5 \times 5)$ reconstruction
(leading to a unit-cell size of 1.81~nm) \cite{Ste89}. A recent
combined x-ray diffraction, STM, and density-functional-theory
study showed that the reconstruction is actually incommensurate
with an approximate $(11 \times 11)$ unit cell [red diamond in
Fig.~\ref{fig1}(a)] \cite{Lun02}. Thus, the Moir\'e pattern
corresponds to an effective $(5.5 \times 5.5)$ unit cell with a
size of 2.00~nm; we observe a size of $1.91 \pm 0.09\, \text{nm}$
[blue dashed diamond in Fig.~\ref{fig1}(a)], in good agreement
with Ref. \onlinecite{Lun02}.

In comparison, Fig.~\ref{fig1}(b) shows STM images of 1~ML
Eu/Gd(0001), with a Moir\'e pattern very similar to that of
Sm(0001). Previously, two different commensurate surface
reconstructions have been reported, $(6 \times 6)$ and $(5 \times
5)$ \cite{Are97,Are98}, which would lead to apparent
Moir\'e-pattern sizes of 2.18~nm and 1.82~nm, respectively
\footnote{Dedkov \emph{et al.} reported a $(4 \times 4)$
reconstruction, supposedly due to a different sample-preparation
procedure, see Ref.\ \onlinecite{Ded06}.}. We find a size of $2.03
\pm 0.06\, \text{nm}$ (blue diamond), very similar to the Sm(0001)
reconstruction. This Moir\'e pattern is not compatible within the
limits of error with either one of the reported reconstructions.
We suggest that the Eu monolayer is also best described by an $(11
\times 11)$ reconstruction (red diamond), with an effective $(5.5
\times 5.5)$ Moir\'e pattern (blue dashed diamond). This
underlines that Sm(0001) and Eu/Gd(0001) are well suited for a
direct comparison, because both their surface reconstructions are
identical and the lattice constants of the hexagonal bulk basal
planes are the same for Sm(0001) and Gd(0001) ($a = 3.63\,
\text{\AA}$) \cite{CRC}.

\begin{figure}
\begin{center}
\includegraphics{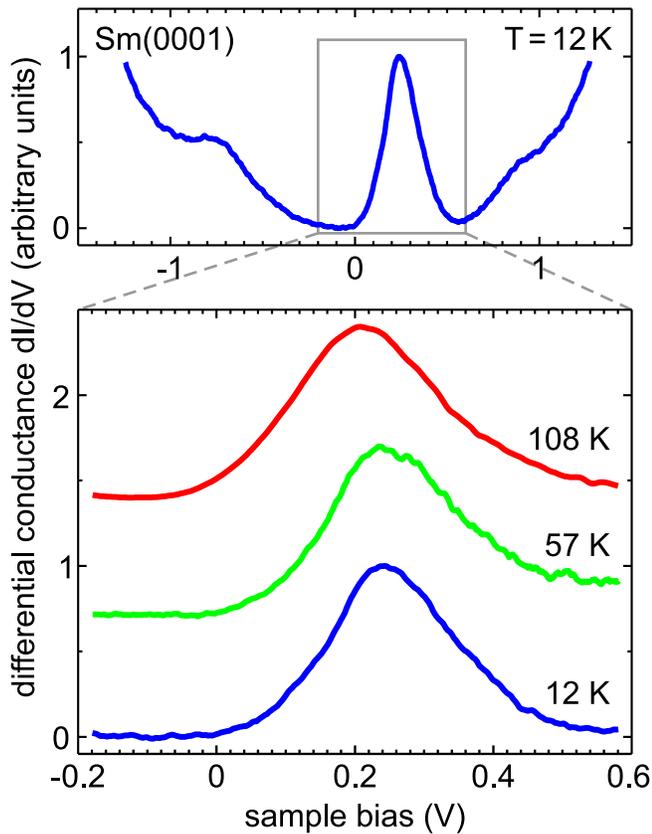}
\caption{\label{fig2}(Color online) STS spectra of Sm(0001): the
large-bias spectrum (top) is dominated by a single peak centered
at 0.25~eV above $E_F$. We interpret this as an unoccupied surface
state of the divalent Sm(0001) surface layer.}
\end{center}
\end{figure}

The STS spectrum of Sm(0001) (Fig.~\ref{fig2}) is dominated by a
narrow resonance at 0.25~meV above $E_F = 0$, independent of the
tip position on the reconstruction pattern. At this energy, all
trivalent Ln metals exhibit a gap in the center of the projected
surface band structure, i.e.\ around the $\overline{\Gamma}$ point
of the Brillouin zone (BZ) \cite{Kur02,Weg06LaLu}. We can
therefore rule out that this strong spectral feature is caused by
a bulk band. Instead, it is reminiscent of the Tamm-like surface
states with $d_{z^2}$ symmetry observed for other Ln-metal
surfaces \cite{Bod98,Bod99,Bau02,Weg06magnon}. We conclude that
the observed peak stems from a surface state of the divalent
Sm(0001) surface. Note that a single peak is observed in the
present case, quite similar to the surface states on non-magnetic
trivalent La(0001) and Lu(0001). Furthermore, STS at various $T$
shows -- apart from the expected slight increase of width with
increasing $T$ -- no significant change of this peak, particularly
no indication of magnetic exchange splitting (see below)
\footnote{The slight shift of the STS spectrum at 108~K is most
likely a $T$-induced artifact of the measurements.}.

\begin{figure}
\begin{center}
\includegraphics{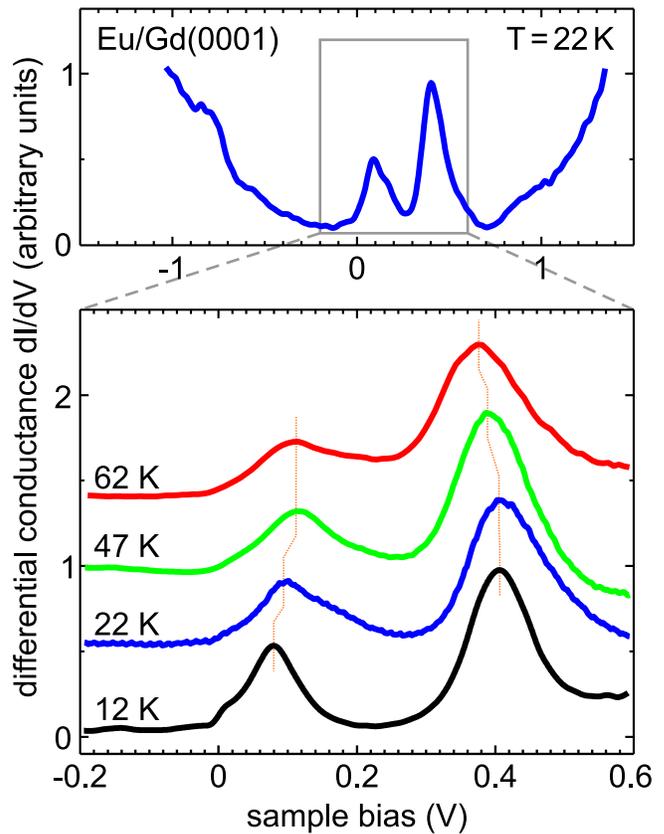}
\caption{\label{fig3}(Color online) STS spectra of 1~ML
Eu/Gd(0001): the large-bias spectrum (top) is dominated by two
peaks at 0.1 eV and 0.4 eV above $E_F$, respectively. STS spectra
at various $T$ (bottom) show that the peak at 0.1~eV shifts to
higher energies with increasing $T$, while the one at 0.4~eV
shifts to lower energies, indicative of an exchange-split surface
state.}
\end{center}
\end{figure}

\begin{figure}
\begin{center}
\includegraphics{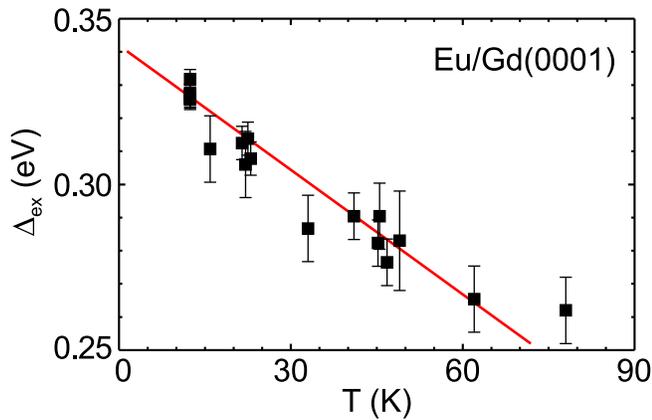}
\caption{\label{fig4}(Color online) Exchange splitting
$\Delta_{\text{ex}}$ of the Eu/Gd(0001) surface state as a
function of $T$, decreasing linearly with $T$ in the studied $T$
range.}
\end{center}
\end{figure}

In contrast, the STS spectrum of 1~ML Eu/Gd(0001)
(Fig.~\ref{fig3}) is dominated by \emph{two} peaks at about 0.1~eV
and 0.4~eV above $E_F$, respectively, again showing no dependence
on tip position. The two peaks are also within the gap in the
center of the projected surface BZ of Gd(0001), reminiscent of the
exchange-split surface states of magnetic trivalent Ln metals
\cite{Weg06magnon}. In order to check this, we studied the
$T$-dependence of the peak positions. With increasing $T$, the
peak at 0.1~eV clearly shifts to higher energies, while the peak
at 0.4~eV shifts down towards $E_F$, reflecting a decrease of the
energy separation of the two peaks with increasing $T$. This
strongly supports an interpretation on the basis of magnetic
exchange splitting into two (majority- and minority-spin)
components separated by $\Delta_\text{ex}$. Fig.~\ref{fig4}
displays $\Delta_\text{ex}$ as a function of $T$. Within the
studied temperature range, $\Delta_\text{ex}$ decreases
approximately linearly with increasing $T$, with a maximum
splitting $\Delta_\text{ex}(T=0) = 342 \pm 2\, \text{meV}$.
Extrapolation towards higher $T$ indicates that the splitting
would vanish at $273 \pm 20\, \text{K}$, which is close to the
$T_C$ of the FM Gd substrate. We note, however, that the exchange
splitting is not expected to decrease linearly with $T$ at higher
$T$, and it might not decrease to zero either \cite{Weg06c}.

To summarize the observations, only one peak can be observed in
the STS spectra of Sm(0001), with no significant $T$ dependence,
whereas the Eu/Gd(0001) surface exhibits two peaks, with a
separation that decreases with increasing $T$. All peaks lie
within a local band gap in the center of the projected surface BZs
of Sm(0001) and Gd(0001), respectively. We therefore interpret
these resonances as signatures of an unoccupied surface state of
the divalent Ln-metal surface layers. Similar to the scenario of
trivalent Ln metals, the narrow, almost Lorentzian-like peak
shapes indicate weak parallel dispersions and hence a relatively
high spatial localization of these surface states
\cite{Bau02,Weg06LaLu}.

The exchange splitting of the Eu/Gd(0001) surface state is caused
by FM coupling of the Eu surface layer to the Gd(0001) substrate.
The maximum exchange splitting of 0.34~eV is about half as large
as that observed for the surface state on pristine Gd(0001),
although both surfaces have a spin $S = 7/2$. This can be
understood by a reduced spin polarization of the Eu layer relative
to the Gd substrate \cite{Are98,Ded06}. Additionally, the $\simeq
20\%$ larger nearest-neighbor distance of Eu-surface atoms should
lead to a smaller interatomic overlap, which should further reduce
the exchange splitting.

In previous studies it was shown that the ground-state exchange
splitting of electronic bands in the Ln metals does not depend on
the total magnetic moment. Instead, it scales almost linearly with
the $4f$ spin \cite{Sch00c,Wes01,Weg06magnon}. From the measured
exchange splitting of Eu/Gd(0001), we can estimate the expected
splitting for $S = 3$ (the spin of divalent Sm) to about 0.29~eV.
The Sm spectra in Fig.~\ref{fig2} do not allow for such a large
splitting. One may argue that the exchange splitting of Sm might
be much smaller, because bulk Sm is AFM. However, within each of
the closed-packed basal planes of the Sm(0001) crystal, the atomic
moments are FM coupled, leading to a local spin polarization of
the subsurface layer \cite{Koe72}. Note that a significant
exchange splitting has also been observed for the surface state on
the complex AFM Nd(0001) \cite{Weg06c}.

We also studied the possibility of describing the STS spectrum of
Sm(0001) by two unresolved components, since the single peak in
the Sm spectrum has a width of $0.22\pm 0.01\, \text{eV}$ (FWHM)
at 12~K, i.e., it is about twice as broad as the peaks in the
Eu/Gd(0001) case. Satisfactory fits could only be obtained for a
peak separation of less than 0.10~eV, i.e.\ less than half the
exchange splitting expected for a magnetic divalent Sm surface. An
argument against magnetic splitting is given by the fact that the
width of the single STS peak increases slightly to $0.27 \pm
0.01\, \text{eV}$ at 108~K by thermal effects, while it should
decrease towards $T_N$ if magnetic splitting would play a role.
The observed increase in peak width is fully compatible with
expectations from both STS data for other Ln-metal surface states
and theory \cite{Reh03,Weg06c,Skr90}. Note also that a doubling of
the width of the STS peak can be understood if we consider that
the spin is not well-defined for the non-magnetic Sm surface state
when an excited electron can scatter into both majority- and
minority-band states of the bulk electronic structure, whereas in
case of Eu/Gd, the (well-defined) spin is conserved. For Sm, this
would double the number of available final states and in turn the
inverse lifetime (proportional to the peak width). All these
considerations support our conclusion that the topmost divalent Sm
surface layer on Sm(0001) is non-magnetic despite the magnetically
ordered trivalent Sm substrate.

The described observations and conclusions should open new
opportunities for atomic-scale local probe studies of mixed-valent
ultra-thin Sm films (see Ref.\ \cite{Nak06}). Through the absence
or presence of a $T$-dependent exchange splitting, it should be
possible to determine the valence state in Sm films. Photoemission
experiments on ultra-thin Sm films revealed such valence changes
as a function of nominal film thickness that might now be analyzed
with high lateral resolution by low-$T$ STM/STS. It is expected
that particularly the combination of high energy resolution with
the high lateral resolution of STS will lead to new insight in the
topic of mixed valency.

%
%
This work was supported by the Deutsche Forschungsgemeinschaft
(DFG), projects KA 564/10-1 and STA 413/3-1. D.W.\ is grateful to
the Alexander von Humboldt Foundation for financial support.

\bibliography{samarium}

\end{document}